\documentclass[11pt,oneside]{article}
\usepackage{a4wide}
\usepackage{mathrsfs}
\usepackage{latexsym,bm}
\usepackage{graphicx}
\usepackage{indentfirst}
\usepackage{amsmath}
\usepackage{amssymb}
\usepackage{color}
\usepackage{hyperref}
\usepackage{epsfig}
\usepackage[titletoc]{appendix}
\usepackage{multirow}
\usepackage{rotating}
\usepackage{epstopdf}
\usepackage{cite}
\usepackage{extarrows}

\newcommand{\be}{\begin{equation}} \newcommand{\ee}{\end{equation}}
\newcommand{\ba}{\begin{array}{c}} \newcommand{\ea}{\end{array}}
\newcommand{\bea}{\begin{eqnarray}} \newcommand{\eea}{\end{eqnarray}}

\newcommand{\order}[1]{\mathcal{O}\left(#1\right)}
\newcommand{\al}{&\!\!\!\!}
\newcommand{\sulr}{{SU$(N)_L\times$SU$(N)_R$}}
\newcommand{\suv}{{SU$(N)_V$}}
\newcommand{\tr}[1]{\left\langle#1\right\rangle}

\newcommand{\email}[1]{\footnote{{\em E-mail address:} \texttt{#1}}}

\begin{document}
\thispagestyle{empty}

\title{
\LARGE \bf
One-loop renormalization of the chiral Lagrangian for spinless matter fields in
the SU({\boldmath$N$}) fundamental representation }

\author{Meng-Lin Du$^{a,}$\email{du@hiskp.uni-bonn.de},
Feng-Kun Guo$^{b,}$\footnote{Emerging leader. {\em E-mail address:} 
\texttt{fkguo@itp.ac.cn}},
Ulf-G.~Mei{\ss}ner$^{a,c,}$\email{meissner@hiskp.uni-bonn.de}\\[2mm]
{\small\it  $^a$Helmholtz-Institut f\"ur Strahlen- und Kernphysik and
Bethe Center for Theoretical Physics,}\\
{\small\it Universit\"at Bonn,  D--53115 Bonn, Germany}
\\
{\small\it  $^b$CAS Key Laboratory of Theoretical Physics, Institute of
Theoretical Physics,}\\
{\small\it  Chinese Academy of Science,  Beijing 100190, China}\\
{\small\it  $^c$Institute for Advanced Simulation, Institut f{\"u}r
Kernphysik and J\"ulich Center for Hadron Physics,}\\
{\small\it Forschungszentrum J{\"u}lich,  D-52425 J{\"u}lich, Germany}
}

\maketitle

\begin{abstract}

We perform the leading one-loop renormalization of the chiral Lagrangian for
spinless matter fields living in the fundamental representation of SU$(N)$.
The Lagrangian can also be applied to any theory with a spontaneous symmetry
breaking of \sulr~to \suv~and spinless matter fields in the fundamental
representation.
For QCD, the matter fields can be kaons or pseudoscalar heavy mesons.  Using the
background field method and heat kernel expansion techniques, the divergences of
the one-loop effective generating functional for correlation functions of single
matter fields are calculated up to $\order{p^3}$. They are absorbed
by counterterms not only from the third order but also from the second order
chiral Lagrangian.

\end{abstract}


\newpage

\section{Introduction}

Chiral perturbation theory (ChPT) provides a systematic and successful approach
for investigating the low-energy behavior of  the Goldstone bosons of the
spontaneous symmetry breaking in
QCD~\cite{Weinberg:1978kz,Gasser:1983yg,Gasser:1984gg}. For quantum
chromodynamics (QCD), the Lagrangian has a chiral symmetry
$\mbox{SU}(N)_L\times\mbox{SU}(N)_R$ in the limit of vanishing quark masses,
where $N$ is the number of flavors under consideration\footnote{Strictly
speaking, the massless QCD Lagrangian has a $\mbox{U}(N)_L\times\mbox{U}(N)_R$
symmetry.
However, the axial U(1)$_A$ is broken at the quantum level due to the anomaly.},
which is spontaneously broken to the vector subgroup SU$(N)_V$. The Goldstone
bosons are provided by the lowest-lying pseudoscalar mesons, to be denoted by
$\phi$.
Since the masses of  the $u,~d$ (and $s$) quarks are small, they can be treated
as a perturbation, and the explicit chiral symmetry breaking due to the quark
masses can be built into ChPT.
In this spirit, the chiral Lagrangian can be constructed order by order in a
double expansion in  the momenta of Goldstone bosons, denoted by $p$, and the
light quark masses with $m_q=\order{p^2}$ (at a fixed ratio of $p^2/m_q$).
The corresponding coefficients are called low-energy constants (LECs) which have
both finite and divergent parts.
To a specific order, both tree level and loop graphs up to that order are
necessary.
Under the proper power counting scheme~\cite{Weinberg:1978kz}, when one
calculates a physical observable, the divergence arising from loops is cancelled
by that of the LECs in the Lagrangian of the proper order, and one gets a finite
and renormalization scale independent result.

Matter fields which are not the Goldstone bosons can be included in ChPT as
well. However, because of the mass of the matter field does not vanish in the
chiral limit, it presents a new energy scale which needs to be treated with
caution. Because of  the introduction of this new energy scale, the naive power
counting is spoiled by the loops containing the matter field
propagators~\cite{Gasser:1987rb}.
Various methods were suggested to restore a consistent power counting. For
the case of matter fields being baryons, there are methods such as the
heavy-baryon ChPT~\cite{Jenkins:1990jv,Bernard:1992qa} and manifestly Lorentz
invariant infrared regularization~\cite{Tang:1996ca,Ellis:1997kc,Becher:1999he}
as well as the extended-on-mass-shell scheme~\cite{Gegelia:1999gf,Fuchs:2003qc}.
In the latter case, since the power counting breaking terms are polynomials of the
external momenta and quark masses, they can be  absorbed into the redefinition
of the LECs. Generally, because of the new energy scale, contrary to the purely
Goldstone boson ChPT, the ultraviolet (UV) divergences generated by the loops
not only shift the higher order LECs, but also the lower order LECs (even the
leading order ones in certain cases).
The same situation happens in the extension of ChPT to include any other matter fields
including ChPT for heavy-flavor
hadrons~\cite{Burdman:1992gh,Wise:1992hn,Yan:1992gz} and SU(2) ChPT for $K\pi$
scattering~\cite{Roessl:1999iu}.

In this paper, we will study the one-loop renormalization of the chiral
Lagrangian for spinless (scalar or peudoscalar) matter fields in the fundamental
representation of SU$(N)$. This kind of theory can be applied to the scattering
between light pseudoscalar mesons and pseudoscalar heavy mesons, SU(2) ChPT
for kaons, and other relevant cases with the same pattern of spontaneous
symmetry breaking and spinless matter fields.

The application of SU(2) ChPT to kaon-pion scattering while treating the kaons
as matter fields ensures a better convergence of chiral expansion than that of
SU(3) ChPT which includes the kaons as Goldstone bosons as
well~\cite{Roessl:1999iu}. This theory is often used in the chiral extrapolation
of lattice results at unphysical up and down quark masses to the physical
values, see, e.g.~Ref.~\cite{Allton:2008pn}.

While there are lots of data for the $K\pi$ scattering, no direct experimental
data for  $D\pi$ scattering exists. Nevertheless, there have been lattice
calculations of the scattering lengths for the scattering of a light
pseudoscalar meson off a charmed
meson~\cite{Liu:2008rza,Liu:2012zya,Mohler:2012na,Mohler:2013rwa,Lang:2014yfa}.
The investigation of the $S$-wave scattering processes in the charmed $D$-meson
sector is important for the understanding of the $0^+$ strange and non-strange
charmed
mesons~\cite{Kolomeitsev:2003ac,Hofmann:2003je,Guo:2006fu,Gamermann:2006nm,
Faessler:2007gv,Lutz:2007sk,Guo:2008gp,Guo:2009ct,Liu:2012zya}. The lattice
results have been used to fix the LECs in the chiral Lagrangian for the $D$
mesons~\cite{Guo:2009ct,Liu:2009uz,Geng:2010vw,Wang:2012bu,Yao:2015qia}.

For purely Goldstone boson ChPT, the systematic one-loop renormalization has
been done in the classical papers by Gasser and
Leutwyler~\cite{Gasser:1983yg,Gasser:1984gg} using the background field method
and  heat kernel expansion techniques. The divergences in the one-loop
effective action are calculated using dimensional regularization which preserves
all the symmetries, and one finds all the counterterms which absorb these
divergences and renormalize the theory at the order $\order{p^4}$.
The complete renormalization of the two flavor heavy baryon chiral pion-nucleon
Lagrangian is performed in Refs.~\cite{Ecker:1994pi,Meissner:1998rw}. Due to the
consistent power counting of HBChPT, the UV divergences of the one-loop
diagrams from the leading order Lagrangian only shift the LECs of the
next-to-next-to-leading order, i.e., $\order{p^3}$.
However, in relativistic baryonic ChPT (or ChPT including heavy mesons or
matter-field kaons), the one-loop diagrams need to be renormalized by LECs from
different orders.

In the present work, we employ the background field method and  heat kernel
techniques to study the renormalization of the chiral Lagrangian for scalar
or pseudoscalar matter fields in the SU$(N)$ fundamental representation to
$\order{p^3}$.\footnote{For the case of scalar/pseudoscalar fields in the
adjoint representations, the one-loop renormalization has been performed in
Ref.~\cite{Rosell:2005ai}.} The paper is organized as follows.
In Section~\ref{lagrangian}, we collect the relevant notations and the
Lagrangians.
In Section~\ref{generatingfunctional}, we derive the relevant one-loop effective
action using the background field method and heat kernel techniques. The
renormalization is carried out and a comparison with purely Goldstone boson
ChPT is also given in Section~\ref{renorm}.
Section~\ref{summary} presents a brief summary.
Finally, we collect the explicit expressions for the ``field strength" tensor
and some useful identities in the Appendices~\ref{app:strength}
and~\ref{app:identities}, respectively, and relevant Lagrangian terms of the
SU(2) ChPT for kaons in Appendix~\ref{app:kaonchpt}.

\section{Chiral Lagrangian of spinless matter fields }
\label{lagrangian}

In this section, we briefly review the chiral generating functional and
Lagrangian describing the interaction between the Goldstone bosons and
spinless matter fields, denoted by $\phi$ and $P$, respectively.
The Goldstone bosons are encoded in an $N\times N$ unimodular, unitary
matrix $U(x)$,
\begin{equation}
U(x)=u^2(x)=\exp{\left(i \frac{\phi}{F_0}\right)},
\end{equation}
where $F_0$ is the pion decay constant in the chiral limit.
$U(x)=u^2(x)=\exp{(i \phi/F_0)}$, and $\phi$ can be expanded in the $N\times N$
traceless Hermitian basis $\phi =\lambda^a \phi^a$ ($\lambda^a$ denotes
the Gell-Mann matrices for $N=3$ and the Pauli matrices for $N=2$) with $\phi^a$
being the corresponding Goldstone boson fields. We
choose the representation in the coset space \sulr/\suv~such that $U(x)$ and
$u(x)$ transform under \sulr~as
\begin{equation}
  U \mapsto g_L^{}\, U\, g_R^\dag,\qquad u \mapsto g_L^{}\, u\, h^\dag = h\, u\,
  g_R^\dag \, ,
\end{equation}
where $g_L\in\,$SU$(N)_L$, $g_R\in\,$SU$(N)_R$, and the compensator field
$h\in\,$SU$(N)$ is a complicated nonlinear function of $g_L,g_R$ and $\phi$
which reduces to the element of the conserved subgroup \suv~when $g_L=g_R$. The
transformation properties of the matter fields are not unique, but it is
convenient to construct the matter fields such that they transform under
\sulr~as
\begin{equation}
 P \mapsto P\, h^\dag, \qquad P^\dag \mapsto h\, P^\dag\, .
\end{equation}
This corresponds to the representation with  $P^\dag$ being in the fundamental representation of \suv.

We will consider the case of the interaction between Goldstone bosons with a
single matter field. The generating functional for the correlation functions of
quark currents between single matter fields is defined via
\be\label{generating}
e^{i Z[j,J,J^\dagger]}=\mathcal{N} \int [d\phi][dP][dP^\dagger]
\exp i \left\{ S_\phi +S_{\phi P} +\int d^4x \left(PJ^\dagger+J P^\dagger\right)
\right\},
\ee
where $S_\phi=\int d^4x \,\mathcal{L}_\phi$ and $S_{\phi P}=\int d^4x\,
\mathcal{L}_{\phi P}$ denote the Goldstone boson and the $\phi P$ chiral
actions, respectively, $J$ and $J^\dagger$ are the sources coupled to the spinless matter fields,
and $j$ denotes various external fields (vector
$v_\mu$, axial $a_\mu$, scalar $s$ and pseudoscalar $p$).
As usual, the quark mass terms will be included in the external scalar source $s$.
The effective chiral Lagrangians can be expanded as
\begin{equation}
  \mathcal{L}_\phi = \sum_{n=1}^\infty \mathcal{L}_\phi^{(2 n)},\qquad
  \mathcal{L}_{\phi P} = \sum_{n=1}^\infty \mathcal{L}_{\phi P}^{(n)}\, ,
\end{equation}
where the  superscripts denote the chiral dimension. We will consider the
renormalization at the leading one-loop order, i.e. $\order{p^3}$ for the
$\phi P$ part and $\order{p^4}$ for the purely Goldstone boson
part.\footnote{One reason to include the purely Goldstone boson part to
$\order{p^4}$ is due to the power counting rule that the Goldstone boson
propagator ${i}/{(q^2-m^2_\phi)}$ is counted as
$\order{p^{-2}}$, while the matter field propagator ${i}/{(q^2-m^2)}$ is
counted as $\order{p^{-1}}$. As a result, the $\order{p^4}$ Goldstone boson
Lagrangian could enter the calculation of the amplitudes for single matter
fields of $\order{p^3}$. One example is given by the contribution of the wave
function renormalization of the Goldstone bosons to the $\phi$-$P$ scattering
amplitudes at $\order{p^3}$~\cite{Yao:2015qia}.}

The leading order Goldstone boson Lagrangian reads
\be
\mathcal{L}_\phi^{(2)}=\frac{F_0^2}{4}\langle u_\mu u^\mu \rangle +\frac{F_0^2}{4} \langle \chi_+ \rangle,
\ee
where $\langle\dots\rangle$ denotes the trace in light-flavor space, and we use
\begin{equation}
u_\mu =i \left(u^\dagger \nabla_\mu^R u-u\nabla_\mu^L u^\dagger\right), \qquad
\chi_{\pm}=u^\dagger
\chi u^\dagger \pm u \chi^\dagger u \, ,
\end{equation}
with the left and right covariant derivatives given by
$\nabla_\mu^R u=\partial_\mu u -i r_\mu u$ and
$\nabla_\mu^L u^\dagger=\partial_\mu u^\dagger-i \ell_\mu u^\dagger$, where
$r_\mu=v_\mu+a_\mu$, $l_\mu=v_\mu-a_\mu$ and $\chi=2 B_0~(s+ip)$ ,with $B_0$ a
positive constant related to the quark condensate. They transform under \sulr~as
\begin{equation}
 u_\mu \mapsto h\, u_\mu\, h^\dag\,, \qquad
 \chi_\pm \mapsto h\, \chi_\pm\, h^\dag\, .
\end{equation}

The chiral effective Lagrangian for spinless matter fields starts from
$\order{p}$, and the leading order terms are given by
\be\label{lagrang1}
\mathcal{L}_{\phi P}^{(1)}=D_\mu P D^\mu P^\dagger -m^2 PP^\dagger,
\ee
where $m$ stands for the mass of the matter fields in the chiral limit, and the
chirally covariant derivative acting on the matter fields are
\begin{equation}
  D_\mu P=\partial_\mu P + P\Gamma_\mu^\dagger\,, \qquad
  D_\mu P^\dagger=\partial_\mu P^\dagger+\Gamma_\mu P^\dagger\, ,
\end{equation}
with the chiral connection $\Gamma_\mu=\frac{1}{2}\left(u^\dagger \nabla_\mu^R
u+u\nabla_\mu^L u^\dagger\right)$. Under \sulr, $D_\mu P$ and $D_\mu P^\dag$
transform in the same way as $P$ and $P^\dag$, respectively.

For later use, we list the $\phi P$ Lagrangian up to
$\order{p^3}$~\cite{Guo:2008gp,Yao:2015qia},
\bea\label{lagrang24}
\mathcal{L}^{(2)}_{\phi P} \al=\al P\left[-h_0\langle\chi_+\rangle-h_1{\chi}_+
+ h_2\langle u_\mu u^\mu\rangle-h_3u_\mu u^\mu\right] {P}^\dag \nonumber\\
\al\al + D_\mu P\left[{h_4}\langle u_\mu
u^\nu\rangle-{h_5}\{u^\mu,u^\nu\}\right] D_\nu {P}^\dag\ , \nonumber\\
\mathcal{L}^{(3)}_{\phi P} \al=\al \bigg[ i~g_1 P[\chi_-,u_\nu]D^\nu
P^\dagger+g_2 P[u^\mu,\nabla_\mu u_\nu+\nabla_\nu u_\mu]D^\nu
P^\dagger
+g_3 P\left[u_\mu,\nabla_\nu u_\rho\right]D^{\mu\nu\rho}P^\dagger
\nonumber\\
\al\al + g_4
P \nabla_\nu \chi_+ D^\nu P^\dagger +g_5 P\langle \nabla_\nu \chi_+\rangle D^\nu
P^\dagger+ h.c.\bigg] \nonumber\\
\al \al + i~\gamma_1 D^\mu P f_{\mu\nu}^+ D^\nu P^\dagger+  \gamma_2
P[u^\mu,f^-_{\mu\nu}]D^\nu P^\dagger~,
\eea
where we have defined
\begin{equation}
D^{\mu\nu\rho}=\{D_\mu, \{D_\nu,D_\rho\}\},
\end{equation}
and
\begin{equation}
f_{\mu\nu}^\pm = u^\dagger \left(\partial_\mu \ell_\nu-\partial_\nu
\ell_\mu-i[\ell_\mu,\ell_\nu]\right)u\pm u\left(\partial_\mu r_\nu-\partial_\nu
r_\mu-i[r_\mu,r_\nu]\right)u^\dagger~.
\end{equation}
The $\gamma_1$ and $\gamma_2$ terms have not been introduced in the literature
to the best of our knowledge. Although they do not contribute to the
scattering, they are necessary for renormalization as one can see later.

For the two-flavor case $N=2$, one may use the Caylay--Hamilton relation, see
Eq.~\eqref{cayleyhamiton2}, to reduce the number of terms. Since the Lagrangian
in this case has been constructed in Ref.~\cite{Roessl:1999iu}, we give the
relations between our LECs and the ones therein ($A_i,B_i$ and $C_i$) (for
completeness, we copy in Appendix~\ref{app:kaonchpt} the relevant Lagrangian
terms from Ref.~\cite{Roessl:1999iu})
\begin{eqnarray}
  \al\al A_1 = 2 h_3 - 4 h_2\,,\quad A_2 = 4(h_5-h_4)\,,\quad A_3 = -h_1\,,\quad
  A_4=-h_0\,, \nonumber\\
  \al\al B_1 = 8 g_2\,,\quad B_3 = -2 g_1\,,\nonumber\\
  \al\al C_3 = 8 g_3, \quad C_5 = -2g_4\,,\quad C_6 = -2g_5\,.
\end{eqnarray}
Notice that we do not have the $B_2$ term here because it is in fact a
$\order{p^4}$ term as can be seen by partial integration. Instead, part of
the $C_3$ term in the $\order{p^4}$ Lagrangian of Ref.~\cite{Roessl:1999iu},
the second line of the $C_3$ term in Eq.~\eqref{eq:LpiK}, is in fact of
$\order{p^3}$, and the relation given above is derived by keeping only that
part in the $C_3$ term.
These corrected assignments can be checked from the explicit expression of the
isospin-3/2 $\pi K$ scattering amplitude in Ref.~\cite{Frink:2002ht}. The $g_5$
and $g_6$ terms are not listed in our original Lagrangian~\cite{Yao:2015qia}
because their contribution to the scattering amplitudes get cancelled by the
wave function renormalization of the matter field at the order $\order{p^3}$.
However, we list them since they are formally of $\order{p^3}$, and they can be
rewritten as the $C_5$ and $C_6$ terms in Ref.~\cite{Roessl:1999iu} by using
the equation of motion for the matter field.

\section{Generating functional to one-loop}\label{generatingfunctional}

In this section, we evaluate the generating functional to the leading
one-loop order using the background field method.
The basic idea of the background field method is to decompose the fields into
classical background fields and quantum fluctuations. After integrating out the
fluctuations, the resulting effective action
actually describes the one-loop contribution of the original action. If
only the divergent parts of the loops are considered, one could employ the heat
kernel techniques to extract the UV divergence of the effective action, which
contains all the possible one-loop divergences and needs to be renormalized by
various counterterms provided by the LECs of the higher order Lagrangians.  To
this end, we perturb the fields $u(x)$ and $P(x)$ around the solutions of the
classical equations of motion $\bar{u}(x)$ and $\bar{P}(x)$,
\bea\label{backgroud}
u^2 \al =\al \bar{u}\,e^{-i \eta}\,\bar u, \nonumber\\
P \al =\al \bar{P}+h,
\eea
where $\eta$ is a traceless Hermitian matrix, $\eta=\eta^a \lambda^a$
($a=1,\ldots,N^2-1$).
Then we substitute the decompositions in Eq.~\eqref{backgroud} into the generating
functional given by Eq.~\eqref{generating}. Since we work up to the leading
one-loop order, we retain only the quadratic terms in $\eta$ and $h$ from
$\mathcal{L}_\phi^{(2)}$ and $\mathcal{L}_{\phi P}^{(1)}$ while
the terms linear in the fluctuations give the equations of
motion, see Appendix~\ref{app:identities}.
For convenience, we collect the fluctuations in the following vectors
\be
\xi_A=
\left(\eta^a,\frac{\sqrt{2}}{F_0}h_i\right)\,,
\quad \xi^\dagger_B=\left(\eta^b,\frac{\sqrt{2}}{F_0}h^\dagger_j\right)^T,
\ee
where $i,j=1,\ldots,N$ while $A$ and $B$ run from 1 to $(N^2-1)+N$ for $N^2-1$
Goldstone boson fluctuations and $N$ matter field fluctuations.
To second order in the fluctuations, the chiral connection $\Gamma_\mu$, the
axial-vector vielbein $u_\mu$ and $\chi_+$ read
\bea\label{umugammamu}
\al \Gamma_\mu \al =\bar{\Gamma}_\mu+\frac{1}{4}\left[\bar{u}_\mu ,\eta\right]
+\frac{1}{8}\left[\eta,\nabla_\mu \eta\right]+\order{\eta^3},\nonumber\\
\al \nabla_\mu \eta \al=\partial_\mu \eta+\left[\bar{\Gamma}_\mu
,\eta\right],\nonumber\\
\al u_\mu \al = \bar{u}_\mu -\nabla_\mu
\eta+\frac{1}{8}\bigl[\eta,[\bar{u}_\mu,\eta]\bigr]+\order{\eta^3},\nonumber\\
\al \chi_+ \al =\bar{\chi}_+ -\frac{i}{2}\{ \bar{\chi}_-,\eta\}
-\frac{1}{8}\bigl\{ \eta, \{\bar{\chi}_+,\eta\}\bigr\}+\order{\eta^3}\, .
\eea
From now on, we will neglect the bars over the classical field configurations
for brevity.
Using the expressions in Eq.~\eqref{umugammamu}, the  terms in the action quadratic in
$\xi$ take the form of
\be
\label{Squad}
S^\text{quad}=-\frac{F_0^2}{2}\int dx ~\xi_A \left( \mathbb{D}_\mu
\mathbb{D}^\mu +\sigma\right)^{AB} \xi^\dagger_B\, .
\ee
Here, the covariant derivative $\mathbb{D}_\mu^{AB}$ is given in matrix form by
\bea
\mathbb{D}_\mu^{AB} =\begin{pmatrix}
    d_\mu^{ab} & \frac{1}{4\sqrt{2}F_0}\big( P[u_\mu, \lambda^a]\big)_j  \\
    \frac{1}{4\sqrt{2}F_0}\big( [u_\mu,\lambda^b]P^\dagger\big)_i &  D_\mu^{ij}
\end{pmatrix},
\eea
where
\begin{eqnarray}
d_\mu^{ab}\al=\al \delta^{ab}\partial_\mu - \frac{1}{2} \tr{
[\lambda^a,\lambda^b]\Gamma_\mu} - \frac{1}{8F_0^2}\left(D_\mu P[\lambda^a,
\lambda^b]P^\dagger- P [\lambda^a,\lambda^b]D_\mu
P^\dagger\right)\,,
\nonumber\\
D_\mu^{ij}\al=\al \delta^{ij}\partial_\mu+(\Gamma_\mu)^{ij}\, ,
\end{eqnarray}
and the non-derivative term $\sigma^{AB}$ stands for
\bea
\sigma^{AB}=\begin{pmatrix}
\sigma_{11}^{ab} & \sigma_{12}^{aj} \\
\sigma_{21}^{ib} & \sigma_{22}^{ij}
\end{pmatrix},
\eea
with
\bea
\sigma_{11}^{ab}\al =\al -\frac{1}{8}\tr{u_\mu
\left[\lambda^a,[u^\mu,\lambda^b]\right]} + \frac{1}{16}\tr{\left\{
\lambda^a,\{\chi_+,\lambda^b\}\right\}}
+\frac{3}{32F_0^2}P [u_\mu,\lambda^a][u^\mu,\lambda^b]P^\dagger
\nonumber\\
\al\al- \frac{1}{64F_0^4}\left( D_\mu P[\lambda^a,\lambda^c]P^\dagger-
P[\lambda^a,\lambda^c]D_\mu P^\dagger\right) \left( D^\mu
P[\lambda^c,\lambda^b]P^\dagger -P[\lambda^c,\lambda^b]D^\mu P^\dagger\right)\,,
\nonumber\\
\sigma_{12}^{aj}\al =\al -\frac{1}{4\sqrt{2}F_0}\big( P[\nabla_\mu
u^\mu,\lambda^a]\big)_j-\frac{3}{4\sqrt{2}F_0}\big( D_\mu
P[u^\mu,\lambda^a]\big)_j\, , \nonumber\\
\al \al + \frac{1}{32\sqrt{2}F_0^3}\left( D_\mu
P[\lambda^a,\lambda^c]P^\dagger-P[\lambda^a,\lambda^c]D_\mu
P^\dagger\right)\big(P[u_\mu,\lambda^c]\big)_j \nonumber\\
\sigma_{21}^{ib}\al =\al \frac{1}{4\sqrt{2}F_0} \left([\nabla_\mu
u^\mu,\lambda^b]P^\dagger\right)_i+\frac{3}{4\sqrt{2}F_0}\left([u^\mu,\lambda^b]
D_\mu P^\dagger\right)_i\, , \nonumber\\
\al \al
+ \frac{1}{32\sqrt{2}F_0^3}\left([u_\mu,\lambda^c]P^\dagger\right)_i \left(D_\mu
P[\lambda^c,\lambda^b]P^\dagger-P[\lambda^c,\lambda^b] D_\mu
P^\dagger\right),\nonumber\\
\sigma_{22}^{ij}\al =\al m^2
\delta^{ij}-\frac{1}{32F_0^2}\left([u_\mu,\lambda^c]P^\dagger\right)_i\big(
P[u^\mu,\lambda^c]\big)_j\,.
\eea
In each of the $\sigma_{11,12,21}$, the last term contains more than two matter
fields, and thus does not contribute to the correlation function of operators
sandwiched between single matter fields. The one-loop term in the
generating functional is a Gaussian integral over the fluctuations $\xi$, which
can be evaluated with standard methods \cite{'tHooft:1973us,donoghue}:
\bea
e^{iZ_\text{loop}}\al =\al \int [d\xi] \exp\left\{-i\frac{F_0^2}{2}\xi^{}_A
\left(\mathbb{D}_\mu \mathbb{D}^\mu+\sigma\right)^{AB}\xi^\dagger_B \right\}
\nonumber\\
\al = \al \mathcal{N} \left(\det [\mathbb{D}_\mu
\mathbb{D}^\mu+\sigma]\right)^{-1/2} \nonumber\\
\al=\al \mathcal{N} \exp \left\{
-\frac{1}{2}\mbox{tr}~\mbox{log}\left(\mathbb{D}_\mu \mathbb{D}^\mu+\sigma\right) \right\}~,
\eea
where $\mathcal{N}$ is an irrelevant normalization, ``tr'' stands for the trace
over the space-time as well as the flavor space spanned by  the basis of the
$\xi_A$. The UV divergences in the generating functional $Z_\text{loop}$ can be
extracted by using the heat kernel expansion (see, e.g., Appendix~B of
Ref.~\cite{donoghue}), and they only show up in the first few expansion
coefficients.
Using dimensional regularization, the relevant terms are
\bea
\label{zloop}
Z_\text{loop}\al =\al \frac{i}{2}\mbox{tr}~\mbox{log}\left(\mathbb{D}_\mu
\mathbb{D}^\mu+\sigma\right) \nonumber\\
\al = \al \frac{1}{2(4\pi)^{d/2}}\int d^4x
\bigg[ \Gamma\left(1-\frac{d}{2}\right)\mu^{d-2} \mbox{Tr}(\sigma)
\nonumber\\ \al \al
+ \mu^{d-4} \Gamma\left(2-\frac{d}{2}\right)\mbox{Tr}\left(
\frac{1}{12}\mathbb{F}_{\mu\nu}\mathbb{F}^{\mu\nu}+\frac{1}{2}\sigma^2 \right)+
\dots \bigg],
\eea
where ``Tr'' denotes the  trace over the space spanned by the $\xi_A$, and the associated
``field strength" tensor
$\mathbb{F}^{AB}_{\mu\nu}=[\mathbb{D}_\mu,\mathbb{D}_\nu]^{AB}$ is given in
Appendix~\ref{app:strength}. Further, $\mu$ denotes the scale of dimensional regularization.
 Only the second term, the $1/(d-4)$ pole,
contributes to the UV divergences at $d=4$.

\section{Renormalization}\label{renorm}

As we are only interested in the UV divergences, we substitute the explicit
expressions of $\mathbb{F}$ and $\sigma$ into Eq.~\eqref{zloop} and
keep only the terms having a $1/(d-4)$ pole.
Using the equations of motion for the classical fields and the identities listed
in Appendix~\ref{app:identities}, one can obtain all possible one-loop
divergences.
Up to $\order{p^3}$, those relevant for single matter fields read
\bea\label{DZloop}
Z_{\phi P}^{\mbox{div}}\al =\al -\frac{\lambda}{F_0^2} \int d^dx \bigg[
\frac{m^2}{24}P\langle u^\mu u_\mu \rangle P^\dagger +\frac{m^2}{24}N~ P u^\mu
u_\mu P^\dagger +\frac{7}{12} D_\mu P \langle u^\mu u^\nu \rangle D_\nu
P^\dagger \nonumber \\
\al \al+\frac{7}{24}N~D_\mu P \{u^\mu,u^\nu\}D_\nu P^\dagger -\frac{3}{64}N\big( P[u^\mu,\nabla_\mu u_\nu+\nabla_\nu u_\mu]D^\nu P^\dagger+h.c.
\big) \nonumber \\
\al \al +\frac{N}{6}i~D^\mu P f^+_{\mu\nu} D^\nu P^\dagger+ \frac{11}{96}N~P[u^\mu,f^-_{\mu\nu}]D^\nu P^\dagger \bigg],
\eea
where $\lambda=\mu^{d-4}[ (4\pi)^{d/2}(d-4)]^{-1}$.

In order to obtain a finite one-loop effective action, the UV divergences in
Eq.~\eqref{zloop} needs to be cancelled by those of the LECs
\be
\mathcal{L}=\sum_i c_i \mathcal{O}_i=\sum_i \left[ c^r_i(\mu) +c^{0}_i
\lambda\right]\mathcal{O}_i\, ,
\ee
where $c_i^r(\mu)$ is the finite part of the $c_i$ and is scale dependent.
From Eq.~\eqref{DZloop} and Eq.~\eqref{puregoldstoneZloop}, it is easy to read
off the divergent parts of the corresponding LECs, which up to $\order{p^3}$ are
given by
\bea\label{hgdiv}
h_{0}^{0}=h_{1}^{0}=0, \quad h_2^{0}=\frac{m^2}{24}, \quad
h_3^{0}=-\frac{m^2}{24} N ,\quad h_4^{0}=\frac{7}{12} , \quad
 h_5^{0}=-\frac{7}{24} N,  \nonumber \\
 g_1^{0}=0, \quad g_2^{0}= -\frac{3}{64} N, \quad
 g_{3,4,5}^0=0,\quad
\gamma_1^{0}=\frac{N}{6}, \quad  \gamma_2^{0}=\frac{11}{96} N~.
\eea
These coefficients determine the scale dependence of the corresponding
renormalized LECs, and the pertinent renormalization group equations read
\begin{equation}
  \frac{\partial c_i^r(\mu)}{\partial \mu} = - \frac{c_i^0}{16\pi^2}\, .
\end{equation}
From Eq.~\eqref{hgdiv}, once clearly sees that the one-loop divergences also
renormalize the LECs at a lower order (here, the $h_i$'s). For $N=3$, the values of
$h_i^0$ and $g_i^0$ agree with those found in an explicit calculation of the
scattering amplitudes~\cite{Yao:2015qia}.

We have checked that integrating out the Goldstone boson fluctuations
leads to the divergence structure of the purely Goldstone boson effective action for
$N$ flavors~\cite{Bijnens:2009qm}:
\bea
\label{puregoldstoneZ}
Z_\phi^{\mbox{div},\phi}
\al = \al -\lambda\int d^dx \bigg[ \frac{N}{48}
\langle u_\mu u_\nu u^\mu u^\nu \rangle + \frac{1}{16} \langle u_\mu u^\mu
\rangle^2 + \frac{1}{8}\langle u_\mu u_\nu \rangle \langle u^\mu u^\nu \rangle
\nonumber\\
\al \al +\frac{N}{24} \langle u_\mu u^\mu u_\nu
u^\nu \rangle + \frac{1}{8}\langle u_\mu u^\mu\rangle \langle \chi_+\rangle +
\frac{N}{8}\langle u_\mu u^\mu \chi_+\rangle +\frac{N^2+2}{16N^2}\langle \chi_+
\rangle^2   \nonumber \\
\al \al +\frac{N^2-4}{16N}\langle \chi_+^2\rangle - \frac{N}{12}i\langle
f^+_{\mu\nu}u^\mu u^\nu\rangle -\frac{N}{24}\langle  f^+_{\mu\nu} f^{+\mu\nu}
\rangle \bigg]~.
\eea
Yet, as can be expected, $Z_\phi^{\mbox{div}}$ also gets contributions due to the
presence of the matter field loops.
The matter fields are expected to be much heavier than the Goldstone
bosons, and therefore their effects in loops on the properties of Goldstone
bosons might be irrelevant, at least for the ChPT of QCD. Nevertheless, they
contribute to the divergence of the generational functional, which is given here
for completeness
\be\label{puregoldstoneZloop}
Z_\phi^{\mbox{div},P}
= -\lambda\int d^dx \bigg[\frac{1}{96}
\langle u_\mu u_\nu u^\mu u^\nu \rangle -\frac{1}{96} \langle u_\mu
u^\mu u_\nu u^\nu \rangle  -\frac{i}{24}\langle f^+_{\mu\nu}u^\mu u^\nu\rangle
-\frac{1}{48}\langle  f^+_{\mu\nu} f^{+\mu\nu} \rangle \bigg].
\ee

\section{Summary}\label{summary}

In this paper, we have performed the renormalization of the chiral Lagrangian
for spinless matter fields living in the fundamental representation of SU$(N)$.
For QCD, the matter fields can be kaons for SU(2) kaon ChPT or pseudoscalar
heavy mesons. Yet, it can also be applied to any other theory with a spontaneous
symmetry breaking of \sulr~to \suv~and spinless matter fields in the fundamental
representation. Using the background field method and heat kernel expansion
techniques, we calculated the divergence of the one-loop effective generating
functional for correlation functions of single matter fields up to the order
$\order{p^3}$, which can be absorbed by various LECs in both the $\order{p^2}$
and $\order{p^3}$ Lagrangians.

\bigskip

\section*{Acknowledgements}

We  thank Bastian Kubis and Akaki Rusetsky for useful discussions.
FKG gratefully acknowledges the hospitality at the HISKP where part of this work
was done. This work is supported in part by DFG and NSFC through funds provided
to the Sino-German CRC 110 ``Symmetries and the Emergence of Structure in QCD''
(NSFC Grant No. 11261130311). FKG is also supported by the Thousand Talents Plan
for Young Professionals. The work of UGM was supported in part by The Chinese
Academy of Sciences  (CAS) President's International Fellowship Initiative
(PIFI) with grant no. 2015VMA076.

\bigskip

\begin{appendix}

\section{The ``field-strength" tensor}\label{app:strength}

The ``field-strength" tensor in Eq.~\eqref{zloop} is
\bea\label{fieldstrength}
\mathbb{F}^{AB}_{\mu\nu}=\begin{pmatrix}
-\frac{1}{2}\langle [\lambda^a,\lambda^b]\mbox{A}_{\mu\nu}\rangle + \Sigma_{11}^{ab} & \Sigma_{12}^{aj} \\
\Sigma_{21}^{ib} & \Gamma_{\mu\nu}^{ij}+\Sigma_{22}^{ij}
\end{pmatrix},
\eea
where~\footnote{Note that $P$ is a $1\times N$ vector, and $P^\dagger$ is a
$N\times 1$ vector. As a result, $P^\dagger P$  is a $N\times N$ matrix.}
\bea
\mbox{A}_{\mu\nu}\al =\al \Gamma_{\mu\nu}+\frac{1}{4F_0^2}\Big(2 D_\mu P^\dagger D_\nu P-2D_\nu P^\dagger D_\mu P + P^\dagger [D_\mu,D_\nu]P-
 [D_\mu,D_\nu]P^\dagger P\Big) \nonumber \\
 \al \al +\frac{1}{(4F_0^2)^2}[P^\dagger D_\mu P-D_\mu P^\dagger P,P^\dagger D_\nu P-D_\nu P^\dagger P],  \nonumber \\
\Sigma_{11}^{ab}\al =\al \frac{1}{32F_0^2}\big( P[u_\mu,\lambda^a][u_\nu,\lambda^b]P^\dagger- P[u_\nu,\lambda^a][u_\mu,\lambda^b]P^\dagger\big), \nonumber \\
\Sigma_{12}^{aj}\al =\al \frac{1}{4\sqrt{2}F_0}\big( D_\mu P[u_\nu,\lambda^a]-D_\nu P[u_\mu,\lambda^a]+P[\nabla_\mu u_\nu-\nabla_\nu u_\mu,\lambda^a]\big)_j \nonumber \\
\al \al -\frac{1}{32\sqrt{2}F_0^3}\Big[ (D_\mu P[\lambda^a,\lambda^c]P^\dagger-P[\lambda^a,\lambda^c]D_\mu P^\dagger )\big(P[u_\nu,\lambda^c]\big)_j
-(\mu \leftrightarrow \nu) \Big],   \nonumber \\
\Sigma_{21}^{ib}\al =\al \frac{1}{4\sqrt{2}F_0}\big( [u_\nu,\lambda^b]D_\mu P^\dagger-[u_\mu,\lambda^b]D_\nu P^\dagger+[\nabla_\mu u_\nu-\nabla_\nu u_\mu, \lambda^b]P^\dagger \big)_i \nonumber \\
\al \al -\frac{1}{32\sqrt{2}F_0^3}\Big[ \big([u_\mu,\lambda^c]P^\dagger\big)_i (D_\nu P[\lambda^c,\lambda^b]P^\dagger-P[\lambda^c,\lambda^b]D_\nu P^\dagger ) -(\mu \leftrightarrow \nu) \Big],  \nonumber \\
\Sigma_{22}^{ij}\al =\al \frac{1}{32F_0^2}\Big[ \big([u_\mu,\lambda^c]P^\dagger\big)_i\big(P[u_\nu,\lambda^c]\big)_j
- \big([u_\nu,\lambda^c]P^\dagger\big)_i\big(P[u_\mu,\lambda^c]\big)_j\Big]~,
\eea
with
\be
\Gamma_{\mu\nu} = [D_\mu,D_\nu]=\partial_\mu \Gamma_\nu -\partial_\nu \Gamma_\mu+[\Gamma_\mu,\Gamma_\nu]~.
\ee

\section{Some useful identities}\label{app:identities}

The equations of motion of the classical fields and some useful identities are
collected here:
\bea
\al\al \nabla_\mu u^\mu =\frac{i}{2}\big( \chi_--\frac{1}{N}\langle \chi_- \rangle \big), \nonumber \\
\al \al D_\mu D^\mu P^\dagger + m^2 P^\dagger =0, \nonumber \\
\al\al \Gamma_{\mu\nu}=[D_\mu,D_\nu]=\frac{1}{4}[u_\mu ,u_\nu]-\frac{i}{2}f^+_{\mu\nu}, \nonumber \\
\al\al \nabla_\mu u_\nu - \nabla_\nu u_\mu =-f^-_{\mu\nu}.
\eea
The Cayley--Hamilton theorem states that every square matrix over a commutative
ring satisfies its own characteristic equation.
For the two-dimensional case, the theorem implies the relation
\be\label{cayleyhamiton2}
\{A,B\}=A\langle B\rangle +B \langle A\rangle + \langle AB \rangle-\langle
A\rangle\langle B\rangle,
\ee
for arbitrary $2\times2$ matrices $A$ and $B$. For the SU(3) case, we have
\be
\langle u^\mu u^\nu u_\mu u_\nu \rangle =-2\langle u^\mu u_\mu u^\nu u_\nu\rangle+\frac{1}{2}\langle u_\mu u^\mu \rangle^2+\langle u^\mu u_\nu \rangle^2.
\ee
Two relations used in deriving Eq.~\eqref{DZloop} are
\bea
\al \al D^\mu P \{ u^\nu,\nabla_\mu u_\nu \} P^\dagger +P \{ u^\nu,\nabla_\mu u_\nu\} D^\mu P^\dagger \nonumber \\
\al=\al D^\mu P \{ u^\nu,\nabla_\mu u_\nu \} P^\dagger - D^\mu P \{ u^\nu,\nabla_\mu u_\nu\}P^\dagger +\order{p^4}\nonumber\\
\al=\al \order{p^4},
\eea
and
\bea
\al \al D_\mu P [u^\mu,u^\nu] D_\nu P^\dagger \nonumber \\
\al=\al -P[\nabla_\mu u^\mu,u^\nu]D_\nu P^\dagger -P[u^\mu,\nabla_\mu u^\nu]D_\nu P^\dagger -P[u^\mu,u^\nu]D_\mu D_\nu P^\dagger \nonumber \\
\al=\al -\frac{1}{2}\big(P[u^\mu,\nabla_\mu u^\nu +\nabla_\nu u^\mu]D_\nu P^\dagger +P[u^\mu,\nabla_\mu u^\nu -\nabla_\nu u^\mu]D_\nu P^\dagger \nonumber\\
\al \al+ P[ u^\mu,u^\nu][D_\mu, D_\nu]P^\dagger+2 P[\nabla_\mu u^\mu,u^\nu]D_\nu P^\dagger \big)\nonumber\\
\al=\al -\frac{1}{2}\Big( P[u^\mu,\nabla_\mu u_\nu+\nabla_\nu u_\mu]D^\nu P^\dagger-P[u^\mu,f^-_{\mu\nu}]D^\nu P^\dagger \nonumber \\
\al \al + i P[\chi_-,u_\nu]D^\nu P^\dagger+\frac{1}{4}P[u^\mu,u^\nu][u_\mu,u_\nu]P^\dagger-\frac{i}{2}P[u^\mu,u^\nu]f^+_{\mu\nu}P^\dagger \Big)\nonumber\\
\al=\al -\frac{1}{2}\Big( P[u^\mu,\nabla_\mu u_\nu+\nabla_\nu u_\mu]D^\nu P^\dagger-P[u^\mu,f^-_{\mu\nu}]D^\nu P^\dagger
 + i P[\chi_-,u_\nu]D^\nu P^\dagger \Big) \nonumber\\
 \al \al +\order{p^4}.
\eea

\section{Relevant Lagrangian terms in kaon SU(2) ChPT}

\label{app:kaonchpt}

Here we list the terms from Ref.~\cite{Roessl:1999iu} relevant for our
comparison:
\begin{eqnarray}
\label{eq:LpiK}
\mathcal{L}_{\pi K}^{(1)} \al =\al D_\mu K^\dag D^\mu K - M_K^2 K^\dag K\, ,
\nonumber\\
\mathcal{L}_{\pi K}^{(2)} \al =\al A_1 \tr{ \Delta_\mu\Delta^\mu
}K^\dag K + A_2 \tr{ \Delta^\mu \Delta^\nu } D_\mu K^\dag D^\nu K + A_3 K^\dag \chi_+ K + A_4
\tr{\chi_+} K^\dag K\, \nonumber\\
\mathcal{L}_{\pi K}^{(3)} \al =\al
B_1 \Bigl( K^{\dagger}\bigl[\Delta^{\nu \mu},
\,\Delta_{\nu}\bigr] D_{\mu}K- D_{\mu}K^{\dagger}\bigl[\Delta^{\nu \mu},
\,\Delta_{\nu}\bigr]K\Bigr) \nonumber\\
\al\al +B_2 \tr{\Delta^{\mu \nu}\Delta^{\rho}}\Bigl( D_{\mu \nu}
K^{\dagger} D_{\rho}K+ D_{\rho}K^{\dagger} D_{\mu \nu}K \Bigr)
\nonumber \\
\al\al +  B_3 \Bigl( K^{\dagger}\bigl[\Delta_{ \mu},\,\chi_-\bigr]
D^{\mu}K- D_{\mu}K^{\dagger}\bigl[\Delta^{ \mu},\,\chi_-\bigr]K\Bigr)\,
,\nonumber\\
\mathcal{L}_{\pi K}^{(4)} \al =\al C_3
\Bigl[\tr{\Delta^{\mu \nu}\Delta^{ \rho}}\bigl( D_{\mu \nu}
K^{\dagger} D_{\rho}K+ D_{\rho}K^{\dagger} D_{\mu \nu}K \bigr) \nonumber\\
\al\al \quad~
- 2\bigl(D^{\mu \nu}K^{\dagger}\Delta_{\mu}\Delta_{\nu \rho}
D^{\rho}K+D^{ \rho}K^{\dagger}\Delta_{\nu \rho}\Delta_{\mu }D^{\mu \nu}
K\bigr)\Bigr]\nonumber \\
\al\al + C_5
\Bigl(D_{\mu}K^{\dagger} \chi_+ D^{\mu}K- M_K^2 K^{\dagger} \chi_+ K \Bigr)
+ C_6 \tr{\chi_+}\Bigl(D_{\mu}K^{\dagger} D^{\mu}K- M_K^2 K^{\dagger}  K \Bigr)
 \nonumber\\ \al\al+ \ldots\,.
\end{eqnarray}
The comparison can be performed by comparing notations in
Ref.~\cite{Roessl:1999iu} with ours:
\begin{eqnarray}
  \Delta_\mu = - \frac{i}{2} u_\mu\, \quad
  \Delta_{\mu\nu} = -\frac{i}{4} \left(\nabla_\mu u_\nu +  \nabla_\nu
  u_\mu\right)\,, \quad
  K^\dag = P\,, \quad K = P^\dag\, ,\quad D_{\mu\nu} = \left\{
  D_\mu,D_\nu\right\}\,.
\end{eqnarray}

\end{appendix}


\end{document}